\documentclass[twocolumn,pre,aps,showpacs]{revtex4}
\usepackage[dvips]{graphicx}

\def\beq{\begin{equation}}
\def\eeq{\end{equation}}
\def\bea{\begin{eqnarray}}
\def\eea{\end{eqnarray}}

\def\pih{{(2\pi\hbar)^3}}

\newcommand{\zeitabl}{\frac{\partial}{\partial t}}

\newcommand{\im}{\mathop{\mathrm{Im}}}
\newcommand{\re}{\mathop{\mathrm{Re}}}
\newcommand{\tm}[1]{{\mathrm{#1}}}

\newcommand{\mvec}[1]{{\mbox{\boldmath$#1$}}}
\newcommand{\qvec}[1]{{\mbox{\boldmath$\overline{#1}$}}}
\newcommand{\quer}[1]{{\overline{#1}}}

\begin{document}

\title{Coupled Mode Effects on Energy Transfer Rates in Two Temperature Plasmas}

\author{J.\ Vorberger}
\affiliation{Centre for Fusion, Space and Astrophysics,
             Department of Physics, University of Warwick,
             Coventry CV4 7AL, United Kingdom}
\author{D.O.\ Gericke}
\affiliation{Centre for Fusion, Space and Astrophysics,
             Department of Physics, University of Warwick,
             Coventry CV4 7AL, United Kingdom}

\date{\today}

\begin{abstract}
We investigate the effects of collective modes on the temperature
relaxation rates in fully coupled electron-ion systems. Firstly,
the well-understood limit of weakly coupled plasmas is considered
and the coupled mode formula within the random phase approximation
is derived starting from the Lenard-Balescu kinetic equation.
We show how the frequency integration can be performed by standard
methods without applying additional approximations. Due to the
well-defined approximation scheme, the results can serve as
a benchmark for more approximate theories and numerical simulations
in this limit. The coupled mode electron-ion transfer rates show
a considerable reduction compared to the Fermi-Golden Rule approach
for certain parameters and very small changes for other systems.
We demonstrate how these coupled mode effects are connected to the
occurance of ion acoustic modes and under which conditions they occur.
Interestingly, coupled mode effects can also occur for plasmas with
very high electron temperatures; a regime, where the Landau-Spitzer
approach is believed to give accurate results. Finally, we extend
the approach to systems with strongly coupled ions by applying static
local field corrections. This extension can substantially increase
the coupled mode effects.
\end{abstract}

\maketitle

\section{Introduction}
Experimental characterization of warm dense matter and strongly coupled plasmas
becomes increasingly powerful with the development of new methods for the
creation and probing of such states. For instance, x-ray scattering allows not
only for the measurement of equation of state data (like in traditional shock
wave experiments, see, e.g., Refs.~\cite{CDCG_98,KHBH_01}), but also to
obtain structural, dynamic, and collective properties of matter
\cite{RWSW_00,RGBD_07,GLNL_07,GGGV_08,KNCD_08}. With these new possibilities,
one is now able to probe the physics of high energy density matter as it is
encountered during inertial confinement fusion or in the interior of planets.

The creation of the these high energy density states in the laboratory requires
large and fast energy inputs into matter. Since static techniques, like diamond
anvil cells, are restricted to lower temperatures and densities by the given
stress point of the material, one relies on dynamic experiments applying
intense particle beams and lasers to heat and compress the material under
investigation. Inevitably, highly nonequilibrium states are produced with the
energy being pumped mainly either into the ion or the electron subsystem. Such
states are of great interest by itself since they reveil many dynamic and
collective properties of the system more clearly.

A good understanding of nonequilibrium states is also needed for definitive
measurements of equilibrium properties. After a very short time of
approximately an inverse plasma frequency, electrons and ions have established
temperatures within their subsystems and a full kinetic description is not
necessary later on. Apart from the final hydrodynamic response, temperature
equilibration thus takes the longest time of all relaxation processes driven
by the intial energy deposition and defines the minimum time delay between the
pump and probe pulses needed for equilibrium measurements.

Temperature equilibration is furthermore interesting since its time scale is
experimentally accessible. Indeed, experiments investigating dense plasmas,
both laser- and shock-produced, found relaxation times considerably longer than
predicted by classical Landau-Spitzer formula \cite{CNXF_92,NCXF_95}. Two
shortcomings of this easy-to-use approach were associated with these
discrepancies: the neglect of collective excitations and the use of classical
collisions. A full quantum binary collision approach yields however even larger
energy transfer rates \cite{GMS_02,GMS_02b,com_1}. Considering independent
collective modes in the electronic and ionic subsystems within the
Fermi-Golden-Rule (FGR) approach yields rates very close to the Landau-Spitzer
results \cite{HZRD_01}. Therefore, collective modes in fully coupled
electron-ion systems seem to be the only candidate to explain the lower
electron-ion energy transfer measured.

More complex descriptions of temperature relaxation involve the interplay of
all contributions to the internal energy. In addition to the kinetic parts
interesting for temperature relaxation, correlations and exchange energies
\cite{GM_05,GMBS_03,GBS_06} and the ionisation equilibrium including
excitations \cite{OBBK_96,BSP_98,GGBS_06} can be important. The latter
processes have been shown to considerable influence the relaxation under
conditions far from equilibrium, but are not sufficient to explain the measured
slower temperature equilibration.

Here, we assume ionisation, recombination, and charge transfer processes to be
completed and concentrate on the electron-ion energy transfer including
collisions and collective excitations in a fully coupled system. We use
a quantum statistical description which avoids any {\em ad hoc} cutoffs known
from classical descriptions and enables us to rigorously derive formulas for
the energy transfer rates within a given approximation scheme.

The occurance and magnitude of coupled mode effects on temperature relaxation
are still under discussion (see Refs.~\cite{DP_98,D_01,DM_07,DM_08,GG_08}). In
the beginning, we therefore consider  weakly coupled plasmas, where the
classical coupling parameter,
$\Gamma_{\! j}  \!=\! Z_{\! j}^2e^2 / a_{\! j} \, k_{\! B} T_{\! j}$ with
$a_{\! j} \!=\! (3 / 4\pi n_{\! j})^{1/3}$ is small for all species. In this
limit, we can employ the well-established random phase approximation (RPA) to
describe the dynamic response of fully coupled systems and the quantum version
of the Lenard-Balescu equation \cite{L_60,B_60,brown} on the kinetic level. We
show that the weak coupling version of the coupled mode formula derived
by Dharma-wardana \& Perrot \cite{DP_98} follows. Moreover, we re-write the
density response functions of their approach in terms of dielectric functions
and obtain a form that can be evaluated by standard integration procedures.

Under certain conditions, we observe a lowering of the energy transfer rates
compared to the Landau-Spitzer and the FGR approaches in the order of a factor
of two. These differences are connected to the occurance of ion acoustic modes
and a related redistribution of weight in the dynamic density response toward
smaller frequencies. This fact also explains why a simpler approximation like
FGR is suitable for other parameters. The most striking fact is that coupled
mode effects can occur also for very high electron temperatures where the
Landau-Spitzer approach is believed to work well.

We have good control over our approximation scheme. The results in the
well-understood weak coupling limit can thus serve as a benchmark for more
approximate approaches. In particular, numerical simulations such as applied in
Refs.~\cite{GGMM_08, JFCC_08} should be tested against these analytic results
since they are based on classical mechanics and apply pseudo-potentials to
approximately incorporate the qantum nature of the electrons as an uncontrolled
approximation.

As an extension to the RPA scheme, we consider strongly coupled ions where the
additional correlation effects are treated on the level of static local field
corrections. The additional shifts in the mode structure can further reduce
the electron-ion energy transfer. For compressed silicon as discussed in
Ref.~\cite{NCXF_95}, we report large reductions although coupled mode effects
on the RPA level do not change the rates much. In addition, we discuss the
influence of degeneracy, mass and ion charge.

\section{Energy Transfer Rates}
\subsection{General Kinetic Description}
Changes of the kinetic energy of species $a$, that is $\dot{E}_a$, are fully
determind by the changes of the one-particle Wigner distribution of the
species $f_a$
\bea
\zeitabl E_a &\!\!=\!\!& \int \frac{d \mvec{p}}{\pih} \,
                              \frac{p^2}{2 m_a} \, \zeitabl f_a(\mvec{p},t)
\nonumber \\
             &\!\!=\!\!& \int \frac{d \mvec{p}}{\pih} \,
                          \frac{p^2}{2 m_a} \, \sum_b I_{ab}(p,t) \,.
\label{balance}
\eea
In the second line, two-particle collision integrals $I_{ab}$ were introduced
by using a general kinetic equation for homogeneous and isotropic systems:
$\partial f_a / \partial t = \sum_b I_{ab}$. The equations for the electron and
ion species are coupled by the collision integrals $I_{ei} = I_{ie}$ and, thus,
the total energy of the plasma is conserved.

There exists a hierachy of approximations for the collision integrals $I_{ab}$
\cite{brown} which via Eq.~(\ref{balance}) also defines the quality of the
related energy transfer rates. The different kinetic equations can be devided
into two classes: the first describes the interactions by binary collisions
only (Landau equation \cite{L_36} and Boltzmann-like equation \cite{D_84}), the
second includes also collective effects such as dynamic screening
(Lenard-Balescu-type kinetic equations \cite{L_60,B_60}). The mutual influence
of electrons and ions via the common dielectric function of the plasma (coupled
collective modes) is here included and predicted to strongly modify the energy
transfer in two-temperature plasmas \cite{DP_98,D_01}.

\subsection{Classical and Quantum Binary Collisions}
The easiest approximation for the two-particle collision integral considers
classical binary collisions. The corresponding energy transfer rates were
derived in the first description of temperature relaxation by Landau
and Spitzer \cite{L_36,S_62}
\beq
\zeitabl E^{LS}_{e \to i} = \frac{3}{2}n_ek_B \frac{T_i-T_e}{\tau_{ei}}
\label{ls_own}
\eeq
with the electron-ion relaxation time
\beq
\tau_{ei} = \frac{3m_em_i}{8\sqrt{2} n_i Z_i^2e^4 \ln\Lambda}
            \left(\frac{k_BT_e}{m_e} + \frac{k_BT_i}{m_i} \right)^{3/2} \,.
\label{ls_time}
\eeq
The plasma properties define the Coulomb logarithm $\ln\Lambda$, where
$\Lambda$ is the ratio of the maximum and minimum impact parameters. Here, we
use a Coulomb logarithm of the form
$\ln\Lambda \!=\! 0.5 \ln(1 + \lambda_e^2/(\varrho_{\perp}^2+\lambda_{dB}^2))$
with the electron screening length
$\lambda_e \!=\! (k_B T_e/4\pi e^2n_e)^{1/2}$, the distance of closest approach
$\varrho_{\perp} \!=\! Z_b e^2 / m_e v_{th}^2$, the deBroglie\- wave length
$\lambda_{dB} \!=\! \hbar / m_e v_{th}$, and the thermal velocity
$v_{th} \!=\! (k_B T_e / m_e)^{1/2}$. This form, which follows by considering
hyperbolic orbits of the electrons, has the advantage to give non-negative
results even for the dense plasmas considered in this paper. For a more
extensive discussion see Ref.~\cite{GMS_02}.

The Landau-Spitzer formula (\ref{ls_time}) suffers from a number of deficits.
The most crucial one is the classical description of electron-ion collisions
which may result in negative Coulomb logarithms and rates. Well-defined energy
transfer rates in the binary collision approximation can however be derived
based on the collision integral of the quantum Boltzmann equation
\cite{zhdan,GMS_02}. If the collision cross sections needed are calculated from
the two-particle Schr\"odinger equation, these rates consider strong
electron-ion collisions as well but collective excitations cannot be included
since statically screened interactions are used. Most binary collision
approaches are also limited to systems with nondegenerate electrons due to the
way the collision cross sections are calculated.

\subsection{Collective Modes in Weakly Coupled Plasmas}
The collective excitations of a weakly coupled plasma are fully accounted for
in the well-known random phase approximation (RPA). Since the electrons and
ions are treated as one combined system in this approach, their mutual
influence is naturally accounted for and, so called, coupled collective modes
arise. This becomes particularly clear when considering the dielectric function
on the RPA level that includes a sum over all species
\beq
\varepsilon^{RPA}(k,\omega) = 1 - \sum_a V_{aa}(k) \, \chi_{aa}^0(k,\omega) \,,
\label{full_dk}
\eeq
where $\chi_{aa}^0$ is the free density response function for the particles of
species $a$.

The kinetic equation corresponding to the RPA is the Lenard-Balescu equation
\cite{L_60,B_60}. Here, we will use its quantum generalization that accounts
for degeneracy \cite{brown}. The energy transfer rates are again obtained by
inserting the appropriate collision integral into the energy
balance (\ref{balance}). Details of the derivation can be found in
Appendix \ref{cm_derivation} and the final result is
\bea
\zeitabl E^{CM}_{e \to i} &\!\!=\!\!&-
            4 \hbar  \sum\limits_i \int \! \frac{d \mvec{k}}{\pih} \!
            \int\limits_0^\infty  \frac{d \omega}{2 \pi} \;
            \omega \, \left| \frac{\tm{V}_{ei}(k)}
                        {\varepsilon^{RPA}(k,\omega)} \right|^2
\nonumber \\
             &&\times
            \im\chi^0_{ee}(k \omega) \, \im\chi^0_{ii}(k \omega) \,
            \Big[ n_B^e(\omega) - n_B^i(\omega)\Big] \,.\nonumber\\
\label{cm_transf}
\eea
$n_B^a(\omega) \!=\! [\exp(\hbar\omega/k_B T_a) - 1]^{-1}$ is a Bose function
that characterises the occupation number of the collective modes and
$V_{ei}(k) \!=\! 4\pi Z e^2/k^2$ is the pure Coulomb potential. The zeros
of the total dielectric function in the denominator determine the collective
excitations of the fully coupled system. Thus, the energy transfer
rate (\ref{cm_transf}) carries the label `CM' for coupled mode. It is
applicable for weakly coupled plasmas without any restriction with respect to
degeneracy. For weakly coupled plasmas, equation (\ref{cm_transf}) is
equivalent to the CM formula given by Dharma-wardana \& Perrot (see Eq.~(50)
in Ref.~\cite{DP_98}) which is demonstrated in Appendix \ref{CM_same}.

We would like to emphasis again that expression (\ref{cm_transf}) represents
a CM formula for the energy transfer rates in weakly coupled, two-temperature
systems. It was derived from the well-established (quantum) Lenard-Balescu
equation in RPA without further approximations. Therefore, it can serve as
a benchmark for more approximate approaches and numerical simulations in the
weak coupling limit.

One of the more approximate, but easier to calculate, approaches for the energy
transfer rates is the Fermi golden rule (FGR) approach. Here, electrons and
ions are treated as separate subsystems. Accordingly, both subsystems have
their own distinct dielectric functions
$\varepsilon_a^{RPA} \!=\! 1 \!-\! V_{aa} \, \chi_{aa}^0$. The expression for
the energy transfer rates reflects this fact by having a product of these
dielectric functions in the denominator \cite{DP_98, G_05}
\bea
\zeitabl E^{FGR}_{e \to i} &\!\!=\!\!&-
            4 \hbar \sum\limits_i \int \!\! \frac{d \mvec{k}}{\pih} \!
            \int\limits_0^\infty  \frac{d \omega}{2 \pi} \;
            \frac{\omega \, \left| \tm{V}_{ei}(k) \right|^2}
                 {\left| \varepsilon_e(k,\omega) \right|^2
          \left| \varepsilon_i(k,\omega) \right|^2}
\nonumber \\
             &&\times
            \im\chi^0_{ee}(k \omega) \, \im\chi^0_{ii}(k \omega) \,
            \Big[ n_B^e(\omega) - n_B^i(\omega)\Big] \,.
\nonumber\\
\label{fgr_transf}
\eea
It can be shown that the FGR rate reduces to Landau-Spitzer (LS) if
classical expressions for the dielectric function are used \cite{HZRD_01,G_05}.

Although both, the CM expression (\ref{cm_transf}) and the
FGR formula (\ref{fgr_transf}), contain effects of collective excitations, they
describe quite different systems since the spectra of the modes do not agree.
To evaluate the corresponding differences in the energy transfer rates is the
main purpose of this paper.

\subsection{Incorporation of Strong Coupling Effects}
In dense plasmas, the Coulomb correlations between the particles can strongly
effect the mode spectrum and scattering cross sections. The extensions towards
strong coupling in the CM formula were already given in the original derivation
by Dharma-wardana \& Perrot \cite{DP_98}.

Strong electron-ion collisions may be accounted for by a weaker electron-ion
pseudo-potential: $\tm{V}_{ei}(k) \to \tm{U}_{ei}(k)$. Spatial correlations,
in particular in the ionic subsystem, can be described by a dielectric function
that includes local field corrections. Here, we will use static local field
correction (SLFC) for the strongly coupled ions only. In this case, the ionic
dielectric function reads \cite{I_82}
\beq
\varepsilon_i(k,\omega) =1-
             \frac{V_{ii}(k) \chi^0_{ii}(k,\omega)}
              {1 + V_{ii}(k) G_{ii}(k) \chi^0_{ii}(k,\omega)}\,.
\eeq
The local field factor $G_{ii}(k)$ is connected to the ionic structure
factor $S_{ii}(k)$ via
\beq
G_{ii}(k) = 1 - \frac{k_BT_i}{n_i V_{ii}(k)}
                \left( \frac{1}{S_{ii}(k)} - 1 \right)\,.
\label{lfc}
\eeq
The spatial correlations in $S_{ii}(k)$ can be calculated by molecular dynamics
or Monte Carlo simulations as well as integral equations like the hypernetted
chain approach \cite{HM_book}.

It should be mentioned here that the changes due to SLFC influence only the
CM rates. Within the FGR approach, the $\omega$-integration is essentially
determined by the f-sum rule \cite{HZRD_01,G_05} which gives the same result
independent of the approximation level used. Therefore, strong coupling effects
are always included in the FGR rates.

\section{Evaluation of the Coupled Mode and FGR Expressions}
\label{eval_cm}
The occurance of collective modes in the systems is connected with zeros in the
dielectric function and results in sharp peaks in the $\omega$-integration.
These peaks prohibit the evaluation of Eqs.~(\ref{cm_transf}) and
(\ref{fgr_transf}) by standard integration routines. We will first show under
which conditions these modes occur and then describe the numerical integration
procedure.

\subsection{Mode Structure in CM and FGR}
\begin{figure}[t]
\includegraphics[scale=0.6,clip=true]{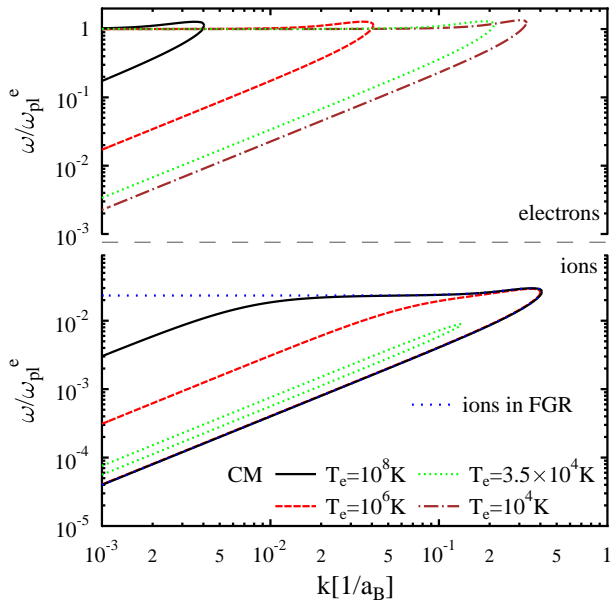}
\caption{(Color online)
         Dispersion relations, $\mbox{Re}\varepsilon(k,\omega) \!=\! 0$,
         for the modes in a hydrogen plasma with an ion temperature of
         $T_i \!=\! 10^4\,$K, a density of $n \!=\! 10^{22}\,$cm$^{-3}$,
         and different electron temperatures. Frequencies are normalized
         to the electron plasma frequency. In the FGR approach, there
         exists an electron and an ion plasmon mode (upper blue dotted
         line) whereas all ion modes become acoustic in the coupled system.}
\label{mode_spectrum}
\end{figure}

The electron modes have the well-known acoustic and plasmon-like branches and
are almost unaffected by the presence of the ions. While the acoustic branch is
strongly damped, the plasmon results in a sharp peak at the plasmon frequency
$\omega_e \!\approx\! (4\pi e^2 n_e/m_e)^{1/2}$ (see Fig.~\ref{mode_spectrum}).
However, these modes are practically not included in the $\omega$-integration
which is effectively limited by the function $\im\chi^0_{ii}(k,\omega)$ to
frequencies less than the ion plasma frequency
$\omega_i \!\approx\! (4\pi Z_i^2 e^2 n_i/m_i)^{1/2}$.

The ionic modes used in the FGR and CM approaches are qualitatively different.
For small momenta, the ion subsystem in FGR has always a plasmon mode which
is similar to the electron plasmon (just scaled by $Z_i$ and $m_i$) as well as
an acoustic mode. The coupling between the electronic and ionic subsystems as
included in the CM formula leads to screening of the ionic plasmon mode which
therefore turns
into a weakly damped upper acoustic branch of the ionic spectrum. The finite
mode frequency at $k \!=\! 0$ is thus shifted to much smaller frequencies that
linearly increase with $k$. Interestingly, this new acoustic branch becomes
more plasmon-like for larger temperature differences and momenta and coincides
with the prediction for an isolated ion system in the unscreened limit with
$T_e \!\to\! \infty$. For the important small $k$ values, the modes are however
for all finite electron temperatures acoustic (in contrast to the FGR modes).

Fig.~\ref{mode_spectrum} also clearly demonstrates that the ionic modes in the
coupled system cease to exist if the temperature difference, $T_e \!-\! T_i$,
becomes too small. The mode frequency is, for small $k$, in very good
approximation given by the zeros of the real part of the dielectric function.
While the product of dielectric functions in the FGR approach has always zeros,
the summation over species in the full dielectric function (\ref{full_dk})
puts strict limits to the occurrence of ionic modes and changes their location
in $\omega$-space; a fact that is crucial for the understanding of the
CM effect on the energy relaxation.

To estimate under which conditions ion acoustic modes in the coupled plasma
occur, we have to balance the electronic terms which can be treated statically
and the frequency-dependent ionic part. It is well-known that a zero in the
real part of the full dielectric function can be achieved under the condition
$T_i \!\ll\! T_e$ \cite{LL_81}. A more precise analysis
(see Appendix \ref{modes_when}) shows that either of the relations
\cite{VG_08}
\bea
T_i \le 0.27\cdot Z_i\,T_e & \qquad \mbox{for}\;\; &n_e\Lambda_e^3 \ll 1\;,
\nonumber\\
T_i \le 0.27\cdot Z_i\,T_F & \qquad \mbox{for}     &n_e\Lambda_e^3 \gg 1
\label{temp_cond}
\eea
must hold to allow for an ion acoustic mode in coupled electron-ion systems.
The upper case gives the limit for nondegenerate electrons while the second
line sets the limit for highly degenerate electrons
($\Lambda_e^2 = 2\pi\hbar / m_e k_B T_e$ is the thermal wave length). In the
high density case, the important temperature scale is thus defined by the
electron Fermi temperature $T_F \!=\! \hbar^2 (3\pi^2 n_e)^{2/3}/2 k_B m_e$.
Accordingly, sharp ion acoustic modes may also occur for $T_e \!<\! T_i$ if the
electron density is high enough.

\subsection{Evaluating the FGR Formula}
The FGR formula for the energy transfer rates (\ref{fgr_transf}) always
features sharp peaks at small wave numbers that are roughly located at the ion
plasmon frequency. The exact value of the mode frequency $\Omega$ can be easily
obtained numerically. To avoid the sharp peaks, these contributions around
the location of the mode are cut out. The cut out contributions can be easily
calculated from the differences to an $\omega$-integral that allows for an
evaluation by the f-sum rule \cite{KSG_90}. Thus, we have to compute
\bea
\lefteqn{\quad
\int\limits_0^\infty \frac{d \omega}{2 \pi} \,
                     \omega \, \Delta N_{ei}(\omega)
                     \frac{\im\varepsilon_e(\omega) \,
                   \im\varepsilon_i(\omega)}
                     {\big| \varepsilon_e(\omega) \big|^2 \,
                      \big| \varepsilon_i(\omega) \big|^2}      }
\nonumber\\
&&\;\,
=\int\limits_0^\infty \frac{d \omega}{2 \pi} \,
                      \omega \, \frac{\im\varepsilon_i(\omega)}
                             {\big| \varepsilon_i(\omega,t) \big|^2} \,
                      F(\omega)
              \left[ 1 - \frac{F(\Omega)}{F(\omega)} \right]
              - \frac{\pi}{2} \omega_i^2 F(\Omega) \,.
\nonumber\\
\eea
Here, $\Delta N_{ei}(\omega) \!=\! n_B^e(\omega) \!-\! n_B^i(\omega)$ is the
difference in the mode occupation numbers. The function $F(\omega)$ includes
the electronic dielectric functions and $\Delta N_{ei}(\omega)$,
that is, $F(\omega) \!=\! \Delta N_{ei}(\omega) \im\varepsilon_e(\omega) /
                          | \varepsilon_e(\omega) |^2$. In this way, the sharp
plasmon peak has been removed and then treated exactly. All remaining integrals
are over smooth functions.

\subsection{Evaluating the CM Expression}
Unfortunately, an evaluation as for the FGR formula is impossible for the
CM integrand. Here, the electronic and ionic parts do not separate and one
cannot expand the integrand into a full dielectric function needed for the
f-sum rule.

The functions under the $\omega$-integral can, however, be rearranged to allow
for an integration with standard routines. For this aim, we artificially
decompose the total dielectric function into an electron and an ion part
\bea
\varepsilon^{RPA}(k,\omega) = 1 \!\!\!& + [\re\varepsilon_e(k,\omega) - 1]
                                \!\!\!& + \im\varepsilon_e(k,\omega) \;\,
\nonumber \\
                                \!\!\!& + [\re\varepsilon_i(k,\omega) - 1]
                                \!\!\!& + \im\varepsilon_i(k,\omega) \,.
\eea
Then, we express the density respose functions in terms of dielectric functions:
$\im\varepsilon_a^{RPA}(k,\omega) \!=\! V_{aa}(k)\im\chi_{aa}^0(k,\omega)$. In
the case of a Coulombic interaction, $V_{ei}^2=V_{ii}V_{ee}$ holds and can be
used to do the transformation above. The $\omega$-integral has now the form
\bea
I_\omega = \int\limits_0^\infty \frac{d \omega}{2 \pi} \,
               \omega \, \Delta N_{ei}(\omega)
               \frac{\im\varepsilon_e(k,\omega) \, \im\varepsilon_i(k \omega)}
                    { \big| \varepsilon^{RPA}(k,\omega) \big|^2}  \,.
\label{coupled_transf_a}
\eea
This form can still have very sharp peaks at the positions of the weakly damped
upper ion acoustic modes. However, these peaks are limited in height since the
real part of the total dielectric functions vanish at the mode frequency and
we have
\bea
\lim_{\re\varepsilon\to 0}
     \frac{\im\varepsilon_e(k,\omega) \, \im\varepsilon_i(k,\omega)}
          {\left| \varepsilon^{RPA}(k,\omega) \right|^2}
\qquad\qquad\qquad\quad
\nonumber \\
   = \frac{\im\varepsilon_e(k,\omega) \, \im\varepsilon_i(k,\omega)}
          {\left| \im\varepsilon_e(k,\omega)
                + \im\varepsilon_i(k,\omega) \right|^2} < 1           \,.
\eea
This rearrangement makes a brute force approach for integrating the CM equation
(\ref{coupled_transf_a}) feasible since extremely narrow peaks have now no
contribution to the integral.

\begin{figure}[t]
\includegraphics[scale=0.6,clip=true]{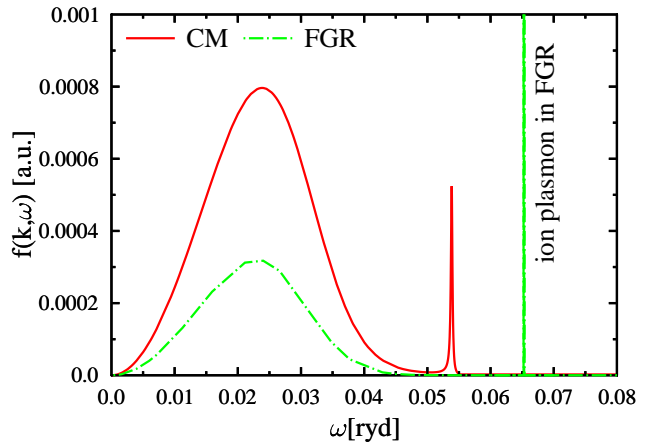}
\caption{(Color online)
         Examples the combination of dielectric functions in the
         $\omega$-integrand of the CM and FGR expressions, {\em i.e.},
         $f_{CM} \!=\! \mbox{Im}\varepsilon_e \mbox{Im}\varepsilon_i /
                                                           |\varepsilon|^2$
         and
         $f_{FGR} \!=\! \mbox{Im}\varepsilon_e\mbox{Im}\varepsilon_i /
                                        |\varepsilon_e|^2|\varepsilon_i|^2$,
         respectively. Considered is a small wave
         number of $k \!=\! a_B^{-1}$ and a hydrogen plasma with
         $T_i \!=\! 10^5\,$K, $T_e \!=\!10^6\,$K, and
         $n \!=\! 10^{24}\,$cm$^{-3}$. Note:\ the height of the ion plasmon
         peak in the FGR approach is 1500 a.u.}
\label{example}
\end{figure}

\subsection{Results of the $\omega$-integration}
Example of the $\omega$-integrands for the two approaches are plotted in
Fig.~\ref{example}. The wave number has been chosen small enough that the
ion acoustic mode exists in the CM approximation. It is clearly visible that
the ion mode is shifted to lower frequencies. Moreover, the height of this
peak has reduced dramatically indicating much larger damping. Although the
particle excitation (left broad peak) is increasing, it cannot compensate the
loss in weight. For a pure ion system, as in the FGR approach, almost the
entire contribution stems from the plasmon peak whereas the main contribution
to the CM integral comes from the particle peak.

\begin{figure}[t]
\includegraphics[scale=0.6,clip=true]{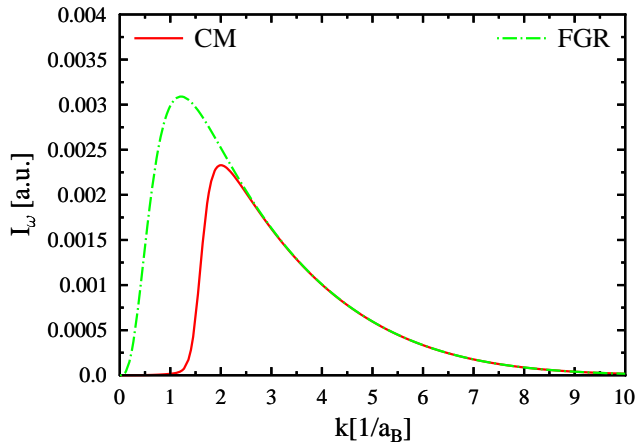}
\caption{(Color online)
         Examples for the $\omega$-integral in the CM and
     FGR expressions for hydrogen as in Fig.~\ref{example}.}
\label{example2}
\end{figure}

The above points are the basics to understand the lowering of the energy
transfer due to coupled collective modes. Figure \ref{example2} demonstrates
that the CM effect is caused by differences at small wave numbers. Here,
the ion acoustic mode exists and the influence of the electrons shifts the
weight to smaller $k$ as shown in Fig.~\ref{mode_spectrum}. Thus, the overall
contributions to the CM integral are of the same order as the
particle peak of the FGR which itself is negligible compared to the ion plasmon
peak. Accordingly, small $k$-values give almost no contribution in the
CM expression compared to the FGR results as shown in Fig.~\ref{example2}. If
these small $k$ are important, significantly reduced CM rates, compared to FGR,
can be expected. For large momentum transfers $k$, the situation is different:
here, the ion acoustic mode (in CM) and the plasmon mode (in FGR) do not exist
anymore and the $\omega$-integrals of both approaches merge. If such larger $k$
dominate in expressions (\ref{cm_transf}) and (\ref{fgr_transf}), no coupled
mode effects can be expected.

\section{Results and Discussion of the Energy Transfer Rates}
We now turn to the analysis of energy transfer rates in dense two-temperature
plasmas applying the CM, the FGR and the LS approaches. We consider a variety
of situations typical for laser- as well as shock-produced plasmas. First, we
will concentrate on weakly coupled plasmas and describe the plasma within the
random phase approximation. Based on the methods presented above, this allows
us to calculate the CM energy transfer rates on a well-established
approximation level (similar to that of the quantum Lenard-Balescu equation)
without any further approximations. In a second step, we will include static
local field corrections to extend the CM approach to plasmas with strongly
coupled ions.

\subsection{Results for Hydrogen within RPA}
\begin{figure}[t]
\includegraphics[scale=0.6,clip=true]{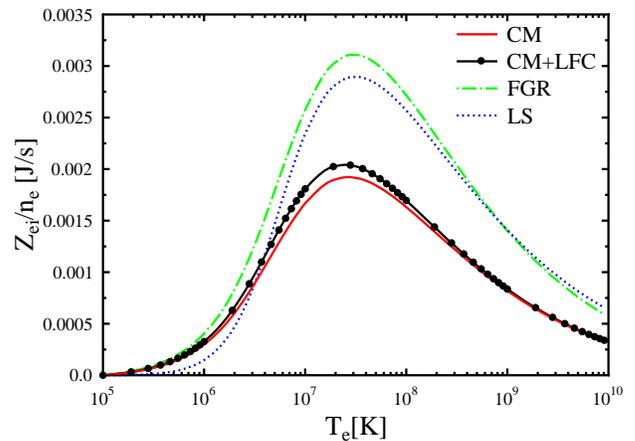}
\caption{(Color online) Electron-ion energy transfer rates in fully
         ionized hydrogen with $n \!=\! 10^{26}\,$cm$^{-3}$ and
     $T_i \!=\! 10^5\,$K versus electron temperature. The applied
     approximations are the coupled mode formula
     within RPA (\ref{cm_transf})
     and with static local field corrections (\ref{lfc}),
     Fermi's golden rule approach (\ref{fgr_transf}), and the
     Landau-Spitzer formual (\ref{ls_own}).}
\label{h_1}
\end{figure}

In a first example, hydrogen at a high density, as needed for the simulation of
inertial confinement fusion scenarios, is considered in Fig.~\ref{h_1}. In this
example, we have relatively warm ions and hotter electrons  (the range of $T_e$
has been extended to see the limiting behavior). All approaches show similar
curves: i) a strong increase of the rates when the electron temperature is just
a few times the ion temperature, ii) maximum of the energy transfer, and
iii) a LS-like asymptotic reduction of the energy transfer $\sim T_e^{1/2}$.
However, considerable quantitative differences arise when comparing the
approximation levels presented.

In the first region, the FGR, CM and CM+LFC results agree since here ion
acoustic or coupled collective modes do not existent. As expected, the
LS formula gives different results in this region since the Coulomb logarith is
not well-defined for these plasmas with relatively cold, degenerate electrons.
At electron temperatures of about $T_e \!=\! 10^6\,$K, ion accoustic modes
start to occur at small $k$ and, accordingly, the CM  results show increasing
deviations from the FGR rate.

Around the maximum of the energy transfer rates the ion acoustic modes are
fully developed. Since they result in smaller $\omega$-integrals for small $k$
(see Fig.~\ref{example2}), the CM results show here a considerable lowering
compared to FGR rates. Although strong coupling effects are small in this
example, the effect of the local field corrections is also largest in this
region.

For very high electron temperatures, the Landau-Spitzer formula is believed to
be accurate due to the very weak electron-ion and electron-electron coupling.
However, our results show that the LS rate only agrees with the FGR rate and
the full binary collision approach (see Ref.~\cite{GMS_02}) in this limit. The
differences between the FGR and LS results for smaller electron temperatures
are due to the different definition of ``Coulomb logarithms'' as the
FGR approach was shown to reduce to a LS-like formula \cite{HZRD_01}. On the
other hand, the CM expression (\ref{cm_transf}) gives considerably reduced
energy transfer rates even in this high electron temperature limit as the ion
accoustic modes are well-pronounced here. CM effects only cease to exist for
high ion temperatures where these modes are not present. Thus CM mode effects
can be also present when describing very hot fusion plasmas.

\begin{figure}[t]
\includegraphics[scale=0.6,clip=true]{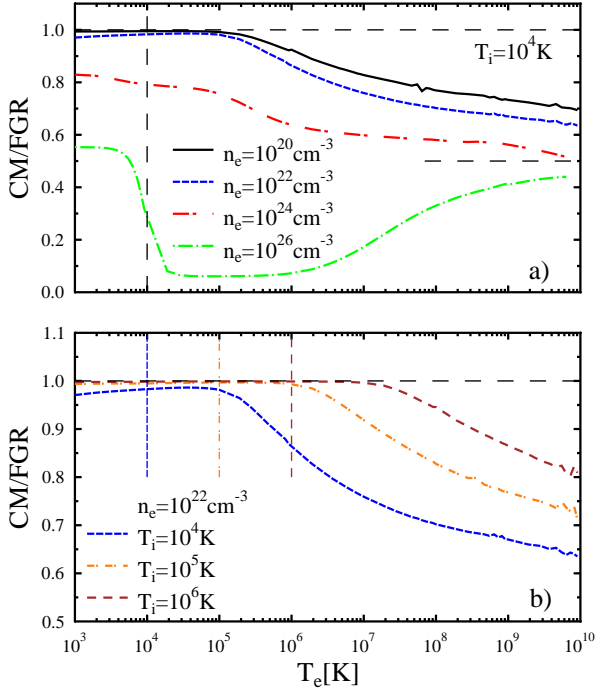}
\caption{(Color online) Ratio of the CM and FGR energy transfer rates
         for fully ionized hydrogen as a function of the electron
     temperature. Panel a) considers various densities while the
     ion temperature is constant; panel b) shows results for
     different ion temperatures at constant density. All results
     are obtained within the random phase approximation. The
     vertical lines mark the condition $T_e \!=\! T_i$.}
\label{cmfgr}
\end{figure}

The relation between different approaches is studied in more detail in
Fig.~\ref{cmfgr}. The first panel demonstrates how an increasing temperature
difference increases the lowering of the energy transfer in the CM compared to
the FGR descriptions. Whereas CM effects slowly develop with increasing
electron temperatures for the lower plasmas densities, they are already present
at $T_e \!=\! T_i$ for densities above $n \!=\! 10^{23}\,$cm$^{-3}$. Here, the
electrons are degenerate with a Fermi temperature that is higher than the ion
temperature and, therefore, ion acoustic modes exist even for similar electron
and ion temperatures (see Eq.~(\ref{temp_cond}) for the occurance of ion
acoustic modes). Under these conditions, the actual temperature difference
becomes irrelevant when searching for CM effects. An extreme case is given by
a density of $n \!=\! 10^{26}\,$cm$^{-3}$ where a lowering of CM rates of up to
90\% can be found near the maximum of the energy transfer.

Of course, an increasing ion temperature has a similar effect as a decreasing
electron density as shown in the second panel of Fig.~\ref{cmfgr} where the well
established CM effect for $T_i=10^4K$ vanishes for higher ion temperatures.
Nevertheless, sufficiently high electron temperatures always guarantee that the
CM rates are lower than FGR or LS rates in these parameter regions.

Interestingly, the ratio of the CM and FGR rates seems to always approach 1/2
in the limit of high electron temperatures. In laser-produced plasmas with
$T_e \!\gg\! T_i$, one should therefore almost always expect reduced energy
transfer rates due to coupled collective modes. On the other hand, the density
in a shock-produced plasma ($T_e \!<\! T_i$) must be high enough to observe
CM effects.

\begin{figure}[t]
\includegraphics[scale=0.6,clip=true]{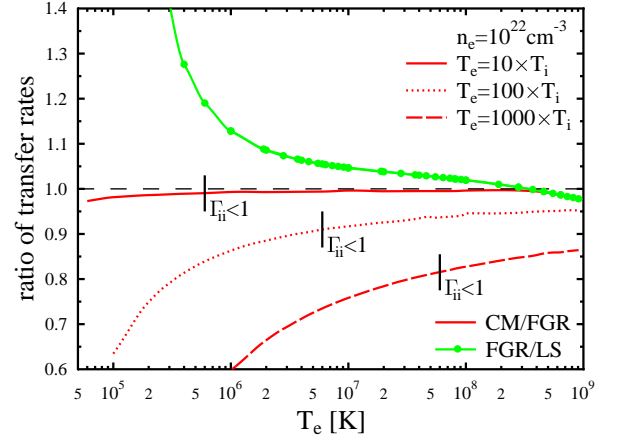}
\caption{(Color online) Ratio of energy transfer rates in
         CM, FGR, and LS in the high temperature (low coupling)
     limit for fully ionized hydrogen. Curves for different
     ratios of electron and ion temperature are given. The
     three curves below unity belong to the ratio of CM to
     FGR rate. The ratio of FGR rate and LS rate is
     independent of the temperature ratio and is given by
     the green dotted line. The region were the ion coupling
     is small is indicated for each case.}
\label{limit}
\end{figure}

Let us analyze the above mentioned high temperature, low density limit of the
CM energy transfer rate in more detail (see Fig.~\ref{limit}). The CM and FGR
approaches do agree in the weak coupling limit {\em if and only if} the
temperature difference between the subsystems is sufficiently low so that no
coupled collective modes can be excited. Larger deviations occur for increased
temperature differences and smaller electron temperatures (more strongly
coupled ions; although RPA is used in Fig.~\ref{limit}). However, even for very
large electron temperatures the CM effects still reduce the coupling between
the electron and ion components.

Moreover, FGR and LS have a ratio independent of the temperature difference. In
addition, numerics show, that even in the weakly coupled case with small
temperature difference, FGR and LS curves intersect rather than converge. This
is due to the poor cut off in the Coulomb integral in LS. Thus the FGR approach
rather than LS formula should be used to compare with experiments or
simulations if one searches for coupled mode effects in the data.

Of particular interest is the question whether the easier FGR is strictly valid
in some cases or not. The difference between FGR and CM originates from the
different denominators. Within RPA, one easily obtains
\bea
|\varepsilon^R(k,\omega)|^2
  \!\!&=&\!\! |\varepsilon_{ee}(k,\omega)|^2 \, |\varepsilon_{ii}(k,\omega)|^2
\nonumber \\
      & & + 2 V_{ee}(k) V_{ii}(k) \, \re\chi^0_{ee}(k,\omega)
                                     \re\chi^0_{ii}(k,\omega)
\nonumber \\
      & & + 2 V_{ee}(k) V_{ii}(k) \, \im\chi^0_{ee}(k,\omega)
                                     \im\chi^0_{ii}(k,\omega)
\nonumber \\
& &-\Big[ |\varepsilon_{ee}(k,\omega)|^2 - 1 \Big]
                           \Big[ |\varepsilon_{ii}(k,\omega)|^2 - 1 \Big] \,.
\label{diff_fgr_cm}
\eea
For the FGR approach to be accurate, the first line only most hold. Strictly,
this is only possible for very large wave numbers $k$. For the interesting
small $k$ values, we must require low plasma densities and the static case
with vanishing imaginary contributions. However, it is essential for the
occurrence of coupled collective modes and the resulting mechanism of energy
transfer to retain the frequency dependence in the dielectric functions.
Whenever, the FGR and CM modes agree (numerics shows us that this requires the
non-occurance of ion acoustic modes), it results from redistributing weights
and is a numerical and approximate agreement of the expressions.

\subsection{Charge State and Strong Coupling Effects}
Coupled mode effects on the energy transfer rates are of course much more
pronounced for plasma ions with higher charges since here the ion acoustic
modes occur for smaller temperature differences or even $T_e \!=\! T_i$
(see Eq.~(\ref{temp_cond}). These modes exist also for higher $k$ values if
the ion charge is higher. Different masses of the ions also shift the modes
to different locations in $k$ space which is, however, irrelevant for the
(relative) CM effect as the modes in FGR behave similarly.

In addition to the charge state, strong coupling effects must be investigated
as the CM reduction of the rates is stronger for colder system as shown in
Fig.~\ref{limit}. In this case, the ions can be strongly coupled and not
described within RPA. Thus, both strong coupling and collective excitations
must be considered in many situations.

\begin{figure}[t]
\includegraphics[scale=0.6,clip=true]{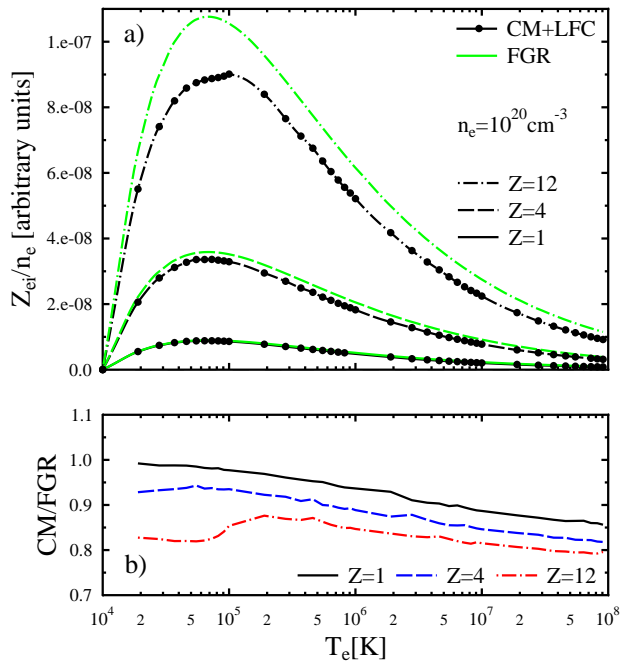}
\caption{(Color online) Effect of the (nonequilibrium) ion
         charge state on the energy transfer rates in different
     approximations for a high--Z plasma.
     Panel a) displays the
     rates and panel b) shows the ratio of the CM and FGR
     results. The ion temperature is $T_i=10^4K$.}
\label{cmfgrbe}
\end{figure}

Figure~\ref{cmfgrbe} shows an example for plasmas with highly charged ions.
When normalized to the electron density the energy transfer rates rise linearly
with the charge $Z$ in the FGR and LS approaches. CM energy transfer rates
naturally depend stronger on the coupled mode structure. Panel b) of
Fig.~\ref{cmfgrbe} shows that the relative reduction for $Z=4$ is higher than
for $Z=1$ so that the charge dependence of CM rates is smaller than $Z$. We
find charge state effects especially for smaller electron temperatures where the
occurance of the ion acoustic mode strongly depends on the charge state of the
ion (see in Eqs.~(\ref{temp_cond})). If the
electron temperature is already high enough to allow for the mode at
$Z \!=\! 1$, the charge state effect is much weaker. For the highest charge
state considered in Fig.~\ref{cmfgrbe}, the influence of the CM rates is even
more strongly pronounced due to strong coupling effects.

\begin{figure}[t]
\includegraphics[scale=0.6,clip=true]{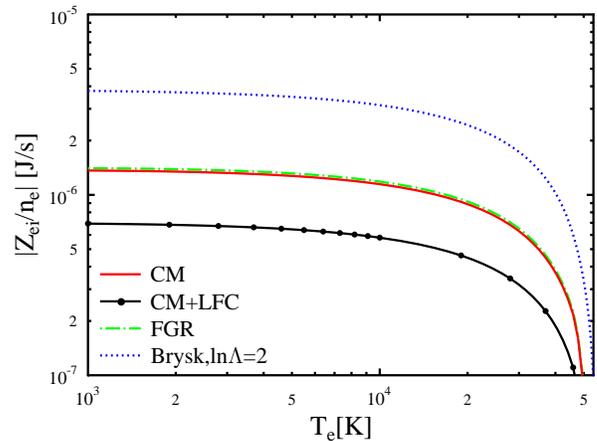}
\caption{(Color online) Energy transfer rate for shocked silicon at
         $n_e \!=\! 5.36 \times 10^{23}\,$cm$^{-3}$, $T_i \!=\! 54539\,$K,
     and with $Z \!=\!4$  in different approximations. The electron
     Fermi temperature is $T_F \!=\! 2.8\times 10^5\,$K. These
     conditions are as in the experiment of Ng {\em et al.}
     \cite{NCXF_95}.}
\label{si}
\end{figure}

The second example in Fig. \ref{si} considers the situation in shocked silicon
plasmas as prepared in the experiment by Ng {\em et al.} \cite{NCXF_95}.
For this case, the electron temperature is lower than the ion temperature. The
electrons are highly degenerate so that according to Eq.~(\ref{temp_cond})
CM effects are possible since the Fermi temperature considerably exceeds the
ion temperature. The ions are also strongly coupled, $\Gamma_{ii} \!=\! 64$,
and local field corrections to the response functions have a larger effect as
in the examples given so far.

When analysing the case, the first point to realise is that neither the usual
Landau-Spitzer approach nor the Brysk formula \cite{B_74} that (partially)
accounts for degeneracy effects is well defined. Both formulas contain
a Coulomb logarithm that either becomes negative or very small when considering
straight line trajectories or hyperbolic orbits, respectively. Thus, the
Coulomb logarithm is usually clamped at some larger value. Here, we use
$\lambda_C \!=\! 2$ for comparison. Still both the LS and Brysk formula give
energy transfer rates that are orders of magnitude larger than the FGR or CM
results which acount for degenerate electrons in a regular fashion. This is
qualitatively consistent with the experimentally observed facts.

Comparing the rates obtained by the CM and FGR expressions, one makes
an interesting observation: on the RPA level both approaches yield more or less
the same results although ion acoustic modes are present in the system. The
weighting due to the electron Fermi distributions favors however larger
$k$ values where no collective mode exist. However, we find a further reduction
of the CM rates when the effect of strong coupling is included by static local
field corrections. Since the FGR approach naturally includes strong coupling
effect \cite{HZRD_01}, no changes occur within this approach. Accordingly, we
find now a factor of roughly two between the rates calculated by the CM and FGR
expressions, respectively. Thus, we can confirm the occurance of coupled mode
effects under these conditions although the main effect arises from the correct
treatment of degeneracy.

\section{Conclusions}
We investigated the electron-ion energy transfer rates in two-temperature
plasmas with special emphasis on coupled mode effects. For weakly coupled
plasmas, the starting point was the quantum version of the Lennard-Balescu
kinetic equation which naturally includes fully coupled collective excitations
in the screening function. It was shown that a straightforward derivation
without further approximations yields a coupled mode expression for the energy
relaxation rates. This kinetic approach is consistent and accurate for weakly
coupled plasmas of any degeneracy. It can be easily extended to plasmas with
strongly coupled ions by including (static) local field corrections.

For certain conditions, the coupled mode expression yield considerably reduced
electron-ion energy transfers when compared to the LS and FGR approaches.
A detailed
analysis showed that this reduction can be traced back to the occurance of ion
acoustic modes and a related redistribution of weight in the dielectric
response function. This reflects the fact that the coupled mode approach
consideres a much more realistic mode structure. Precise conditions for the
occurrence of ion accoustic modes and the coupled mode reduction of the energy
transfer rates were derived. Interestingly, the CM reduction is preserved for
very high electron temperatures where the rates are roughly a factor of two
less than the ones calculated with the FGR or LS approaches. An agreement
between CM and FGR expressions can only be reached if the temperature
difference is small or the ion temperature is high enough to prohibit the
occurance of ion acoustic modes.

The results presented here are concistent with experimental
\cite{RWSW_00,CNXF_92,NCXF_95} and theoretical \cite{DP_98,D_01} findings
of lowered electron-ion energy transfers and thus longer equilibration times.
In particular, the strong reduction of the equilibration rates in shocked
silicon is in qualitative agreement with measurements. The experiments probed
however regimes where collective modes as well as strong binary collisions and
changing potential energies are of importance. Thus, a full description must
also include these effects at least on a level as described in
Refs.~\cite{GMS_02,GM_05,G_05}. The conditions for the occurance of CM effects
derived also explain why no or very small CM reductions have been seen in
recent numerical simulations \cite{GGMM_08,DD_08,JFCC_08}.

\section*{Acknowledgement}
We gratefully acknowledge financial support from the Engineering and Physical
Sciences Research Council and stimulating discussions with Prof. W.-D. Kraeft.

\begin{appendix}

\section{Derivation Coupled Mode Expression in Random Phase Approximation}
\label{cm_derivation}
We start with the quantum Lennard-Balescu equation for the electron distribution
$f_e(p,t)$ in homogeneous and isotropic plasmas
\cite{brown}
\begin{widetext}
\bea
\zeitabl f_e(p,t) &\!\!=\!\!&
              \frac{1}{\hbar} \sum\limits_i
              \int \frac{d \mvec{p}^\prime}{\pih} \frac{d \qvec{p}}{\pih}
              \frac{d \qvec{p}^\prime}{\pih} \,
              \left| \frac{\tm{V}_{ei}(\mvec{p} - \qvec{p})}
              {\varepsilon^{RPA}\left( \mvec{p} - \qvec{p},
                                 E_e(p) - E_e(\quer{p}),t \right)}\right|^2
\nonumber \\
             &\!\!~\!\!& \times
              2\pi \delta\!\;\Big( E_e(\mvec{p}) + E_i(\mvec{p}^\prime)
                                  - E_e(\qvec{p}) - E_i(\qvec{p}^\prime) \Big)
              \;\pih\, \delta\!\!\;\Big( \mvec{p} + \mvec{p}^\prime
                                      - \qvec{p} - \qvec{p}^\prime \Big)
\nonumber \\
             &\!\!~\!\!& \times
             \Big\{ f_e(\qvec{p},t) f_i(\qvec{p}^\prime,t)
                     \, [1 - f_e(\mvec{p},t)] \, [1 - f_i(\mvec{p}^\prime,t)]
- f_e(\mvec{p},t) f_i(\mvec{p}^\prime,t)
                     \, [1 - f_e(\qvec{p},t)] \, [1 - f_i(\qvec{p}^\prime,t)]
             \Big\} \,.
\nonumber \\
\label{balescu_gl}
\eea
\end{widetext}
The sum runs over all species, but only the ions contribute to the energy
relaxation. $E_a(p) \!=\! p^2/2 m_a$ denotes the kinetic energy of particles of
species $a$ and $V_{ab}(k) \!=\! 4\pi Z_a Z_b e^2/k^2$ the (unscreened) Coulomb
potential. Dynamic screening is described by the retarded dielectric function
which is used here in RPA
\bea
\varepsilon^{RPA}(\mvec{p},E,t) &=&
           1 - \sum_a \; V_{aa}(p) \, \chi^0_{aa}(\mvec{p},E,t) \,.
\label{dielec}
\eea
The upper expression relates the screening function to the density response
function of free particles given by
\beq
\chi^0_{aa}(\mvec{p},E,t) = \int \frac{d \mvec{p}^\prime}{\pih}
      \frac{f_a(\mvec{p}^\prime,t) - f_a(\mvec{p}^\prime + \mvec{p},t)}
           {E + E_a(\mvec{p}^\prime) - E_a(\mvec{p}^\prime + \mvec{p})
            + i \epsilon} \,,
\label{density_response}
\eeq
which is, in turn, determined by the electron or the ion distribution function.

The electron-ion energy transfer rates are now obtained by multiplying the
Lenard-Balescu equation (\ref{balescu_gl}) by the electron energy,
$E_e=p^2/2m_e$, and an integration over the free momentum (see balance equation
(\ref{balance}).

For the further proceedings, it is useful to use the momentum and the energy
transfer during the collision, $\mvec{k} \!=\! \mvec{p} \!-\! \qvec{p}$ and
$\omega \!=\! E_e(p) \!-\! E_e(\quer{p})$, respectively. With these new
variables, one can now apply the relations between Fermi-distributions of the
plasma species and Bose functions
$n_B^a(\omega) \!=\! [\exp(\hbar\omega/k_B T_a) \!-\! 1]^{-1}$, namely
\bea
\lefteqn{\!\!\!\!\!f_a(\mvec{p}) \, [1 - f_a(\mvec{p} + \mvec{k})] =}
\nonumber \\
 &&  [f_a(\mvec{p} + \mvec{k}) - f_a(\mvec{p})] \,
               n_B^a\!\Big(E_a(\mvec{p}) - E_a(\mvec{p} + \mvec{k}) \Big) \,,
\\
\lefteqn{\!\!\!\!\!f_a(\mvec{p} + \mvec{k}) \, [1 - f_a(\mvec{p})] =}
\nonumber \\
 &&   [f_a(\mvec{p}) - f_a(\mvec{p} + \mvec{k})]
               n_B^a\!\Big(E_a(\mvec{p} + \mvec{k}) - E_a(\mvec{p}) \Big) \,.
\eea
The set of distributions in the third line of the collision integral of the
Lenard-Balescu equation (\ref{balescu_gl}) can then be written as
\bea
\Big\{ f() \Big\} \!\!&=&\!\!
[f_e(\qvec{p} + \mvec{k}) - f_e(\qvec{p})] \,
                    [f_i(\mvec{p}^\prime) - f_i(\mvec{p}^\prime + \mvec{k})]
\nonumber\\
               \!\!& &\!\! \times \Big[ n_B^e(-\omega) \, n_B^i(\omega)
                           - n_B^e(\omega)\, n_B^i(-\omega) \Big] \,.
\eea
With these transformations, we obtain for the electron-ion energy transfer rate
\begin{widetext}
\bea
\zeitabl E^{CM}_{e \to i} &\!\!=\!\!&
             - \frac{1}{\hbar}  \sum\limits_i
            \int \frac{d \qvec{p}}{\pih} \frac{d \mvec{k}}{\pih}
            \frac{d \omega}{2 \pi} \;
            E_e(\qvec{p} + \mvec{k}) \;
            \left| \frac{\tm{V}_{ei}(\mvec{k})}
                        {\varepsilon^R(\mvec{k} \omega)} \right|^2
            \Big[ f_e(\qvec{p} + \mvec{k}) - f_e(\qvec{p}) \Big]
        2 \pi\delta\!\!\;\Big( \omega - E_e(\qvec{p} + \mvec{k}) +
                                       E_e(\qvec{p}) \Big)
\nonumber \\
               &\!\! \!\!& \times
               \Big[ n_B^e(-\omega) \, n_B^i(\omega)
                           - n_B^e(\omega)\, n_B^i(-\omega) \Big] \;
           \int \frac{d \mvec{p}^\prime}{\pih} \:
           [f_i(\mvec{p}^\prime) - f_i(\mvec{p}^\prime + \mvec{k})] \:\,
            2 \pi \delta\!\!\;\Big( \omega - E_i(\mvec{p}^\prime
                                  + \mvec{k}) + E_i(\qvec{p}^\prime) \Big) \,.
\nonumber \\
\eea
\end{widetext}
The integral in the second line of this expression is just the definition of
the imaginary part of the free particle density response function for the ionic
subsystem $\im\chi^0_{ii}(k \omega)$ as defined by (\ref{density_response}).
The difference of Bose functions in the second line determines the direction of
the energy transfer. After another variable transformation to
$\omega^\prime \!=\! -\omega$, $\mvec{k}^\prime \!=\! - \mvec{k}$, and
$\mvec{p}^\prime \!=\! \qvec{p} \!-\! \mvec{k}^\prime$ (changing the sign) in
the second term proportional to $n_B^e(\omega)\, n_B^i(-\omega)$, this second
term has the same form as the first one, except that the energy in front of the
screened potential is $E_e(\qvec{p})$ instead of $E_e(\qvec{p} \!-\! \mvec{k})$.
Applying the energy conserving $\delta$-function in the collision integral
yields
\beq
E_e(\qvec{p} + \mvec{k}) - E_e(\qvec{p}) = \omega \,.
\eeq
Now the remaining electron distributions together with the energy conserving
$\delta$-function give also the definition of an imaginary part of the free
density response function; this time, the one for the electrons,
i.e.\ $\im\chi^0_{ee}(k \omega)$. The energy transfer rate is thus given by
\bea
\zeitabl E^{CM}_{e \to i} &\!\!=\!\!& -
            4 \hbar  \sum\limits_i \int \frac{d \mvec{k}}{\pih}
            \int\limits_{-\infty}^\infty\!\!  \frac{d \omega}{2 \pi} \,
            \omega \, \left| \frac{\tm{V}_{ei}(k)}
                        {\varepsilon^{RPA}(k \omega)} \right|^2 \;
\nonumber \\
             &\!\!  \!\!& \times
            \im\chi^0_{ee}(k \omega) \, \im\chi^0_{ii}(k \omega) \,
            n_B^e(-\omega) \, n_B^i(\omega) \,.\nonumber\\
\eea
If we use the fact that $[1/2 \!-\! n_B(\omega)]$ and $\im\chi^0_{aa}(\omega)$
are odd functions with respect to $\omega$, we can rearrange the upper
expression in the form
\bea
\zeitabl E^{CM}_{e \to i} &\!\!=\!\!& -
            4 \hbar  \sum\limits_i \int \frac{d \mvec{k}}{\pih}
            \int\limits_0^\infty\!\!  \frac{d \omega}{2 \pi} \;\:
            \omega \, \left| \frac{\tm{V}_{ei}(k)}
                        {\varepsilon^{RPA}(k \omega)} \right|^2 \;
\nonumber \\
             &\!\!   \!\!& \times
            \im\chi^0_{ee}(k \omega) \, \im\chi^0_{ii}(k \omega) \,
            \Big[ n_B^e(\omega) - n_B^i(\omega)\Big] \,.
\nonumber\\
\label{coupled_transf_a_appen}
\eea
This expression gives the electron-ion energy transfer rate including the
effects of fully coupled collective modes in the electron-ion system. Similar
to the Lenard-Balescu equation that was used as a starting point, it describes
weakly coupled plasmas without restriction with respect to the degeneracy.

\section{Equivalence to Coupled Mode Description by Dharma-wardana \& Perrot}
\label{CM_same}
The coupled mode expression derived by Dharma-wardana \& Perrot, equation (50)
in Ref.~\cite{DP_98}, reads
\bea
\dot{E}_{rlx }&\!\!=\!\!&
        \hbar\int\limits_0^{\infty}\frac{d\omega}{2\pi}
             \int \frac{d{\bf k}}{(2\pi\hbar)^3} \; \omega\, V_{ie}^2(k)
\nonumber\\
             &\!\!  \!\!& \times
        \frac{\Delta N_{ei}(\omega) \, A^i(k,\omega) \,A^e(k,\omega)}
             {|1-V_{ie}^2(k)\,\chi_{ee}(k,\omega) \, \chi_{ii}(k,\omega)|^2}\,,
\label{coupled_transf_dp}
\eea
where we assumed all potentials to be of Coulomb type for simplicity. The
difference of Bose functions is here the same as in
Eq.~(\ref{coupled_transf_a_appen}):
$\Delta N_{ei} \!=\!n_B^e(\omega) - n_B^i(\omega)$. The functions
$A^a(k,\omega)$ are related to the imaginary parts of the response functions
$\chi_{aa}(k,\omega)$
\beq
A^a(k,\omega) = - 2\mbox{Im}\chi_{aa}(k,\omega) \,.
\eeq
One should however notice that these are full density response functions of a
coupled system.

The density response function $\chi_{aa}$ for a single species can also be
written in terms of the polarization function $\Pi_{aa}$ and the dielectric
function $\varepsilon$ of the (fully coupled) medium. Neglecting cross terms,
we have \cite{green}
\beq
\chi_{aa}(k,\omega)
=\frac{\Pi_{aa}(k,\omega)}{1-V_{aa}(k)\Pi_{aa}(k,\omega)}
=\frac{\Pi_{aa}(k,\omega)}{\varepsilon_{aa}(k,\omega)} \,.
\eeq
This is nothing else than the equation of motion for the density response
function. Real and imaginary parts of the response function can then be
expressed as
\bea
\mbox{Re}\chi_{aa}(k,\omega) &\!\!=\!\!&
\frac{\mbox{Re}\varepsilon_a(k,\omega)-|\varepsilon_a(k,\omega)|^2}
{V_{aa}(k)| \varepsilon_a(k,\omega)|^2} \,,
\\
\mbox{Im}\chi_{aa}(k,\omega) &\!\!=\!\!& -
\frac{\mbox{Im}\varepsilon_a(k,\omega)}
{V_{aa}(k)|\varepsilon_a(k,\omega)|^2}\,.
\label{imchi}
\eea
With the help of these expressions, the denominator in
expression (\ref{coupled_transf_dp}) which constitudes the differences to the
FGR formula becomes
\bea
\lefteqn{|1-V_{ie}^2(k)\,\chi_{ee}(k,\omega)\chi_{ii}(k,\omega)|^2}
\nonumber\\
&& \qquad\qquad\qquad = \frac{|\varepsilon(k,\omega)|^2}
{|\varepsilon_e(k,\omega)|^2|\varepsilon_i(k,\omega)|^2} \,.
\label{new_exp}
\eea
The dielectric function in the numerator of Eq.~(\ref{new_exp}) is the one for
the full system given by  $\varepsilon=1-\sum_a V_{aa}\Pi_{aa}$. The partial
dielectric functions $\varepsilon_a$ in this expression are cancled by the one
contained in the imaginary part of the full density response function
$\chi_{aa}$ when the latter are written as in eq.~(\ref{imchi}). Accordingly,
one obtains an expression identical to Eq.~(\ref{coupled_transf_a_appen}) when
inserting these definitions and transformations into
Eq.~(\ref{coupled_transf_dp}).

\section{Conditions for the Occurance of Ion Acoustic Modes}
\label{modes_when}
It is well known that ion accoustic modes occur in non-equilibrium plasmas with
$T_e \!\gg\! T_i$ \cite{LL_81}. Here, we give a more precise estimate of the
necessary temperature relation needed to allow for this type of collective
modes in weakly coupled plasmas.

For ion acoustic modes to exist, the real part of the total dielectric function
must vanish. We first consider the ionic contribution in non-degenerate limit
which is almost always sufficient for the heavy ions. In RPA, the polarisation
function is identical to the free particle density response function, that is
$\Pi_{ii} \!=\! \chi_{ii}^0$. Its real part can be written in terms of the
confluent hypergeometric function \cite{green,brown}
\beq
\mbox{Re}\chi^0_{ii}(k,\omega) =
  \frac{n_i}{k_BT}  \left[ 1 - \frac{\omega^2 m_i}{p^2k_BT}  \,
            _1F_1\!\left(1,\frac{3}{2},
                     -\frac{\omega^2 m_i}{2p^2k_BT}\right) \right]
\label{reepsii}
\eeq
which is defined over an integral \cite{grad}. However, one can approximate
this function for the special first two arguments needed by the following
Pad\'e formula
\cite{brown}
\bea
\lefteqn{_1F_1\!\left(1,\frac{3}{2},-x \right) = }
\nonumber\\
&&       \frac{1 + \frac{x}{3} + \frac{x^2}{10} + \frac{x^3}{42}
               + \frac{x^4}{218} + \frac{7x^5+x^6}{9360}}
            {1 + x + \frac{x^2}{2} + \frac{x^3}{6} + \frac{x^4}{24}
               + \frac{x^5}{120} + \frac{x^6}{720} + \frac{x^7}{4860}} \,.
\label{einsfeins}
\eea
As a required condition for ion acoustic modes to occur, the minimum of the
response functions (\ref{reepsii}) must at least compensate the electronic
contribution plus unity. We thus search for an approximation of
Eq.~(\ref{einsfeins}) that conserves this minimum. Performing a frequency
derivative, we find its location to be approximately at
$\omega_0 \!=\! 2.36 \, k \, (k_B T_i /m_i)^{1/2}$. Using the minimum frequency
$\omega_0$ in Eq.~(\ref{reepsii}) with the hypergeometric function given by
Eq.~(\ref{einsfeins}) yields
\begin{equation}
\mbox{Re}\;\!\varepsilon_i(k,\omega_0) = 1 - 0.27 \, \frac{\kappa_i^2}{k^2} \,,
\label{ion_min}
\end{equation}
where $\kappa_i=(4 \pi Z_i^2 e^2 n_i/k_B T_i)^{1/2}$ is the inverse of the
ion part of the classical Debye screening length.

The electronic part of the dielectric function can be very well estimated in
the zero frequency (static) long wave length limit
\begin{equation}
\mbox{Re}\;\!\varepsilon_e(k,0) = 1 + \frac{\kappa_e^2}{k^2} \,.
\label{electron_static}
\end{equation}
Again, $\kappa$ denotes the inverse screening length. For small momenta, the
unity in Eqs.~(\ref{ion_min}) and (\ref{electron_static}) can be neglected. The
condition for the ion acoustic modes to emerge becomes then
\begin{equation}
\kappa_e^2 \le 0.27 \,\kappa_i^2 \,.
\label{kappa_result}
\end{equation}
Eq.~(\ref{kappa_result}) quantifies the long known condition of $T_i\ll T_e$
\cite{LL_81} for the ion acoustic waves to occur. Direct relations for the
electron and ion temperatures needed may be obtained by inserting
an appropriate expression for the electron screening length. In the
nondegenerate limit, it is given by the electron Debye length and one obtains
the upper relation of Eq.~(\ref{temp_cond}). For highly degenerate electrons,
Thomas-Fermi screening with
$\kappa_e \!=\! \kappa_{TF}^2 \!=\! 4m_e e^2(3n_e/\pi)^{1/3}$ must be used.
Since the Thomas-Fermi length is temperature independent, the role of the
electron temperature is taken over by the electron Fermi temperature which
yields the second condition in Eq.~(\ref{temp_cond}) for the existence of the
ion acoustic modes. Accordingly, they can also occur in degenerate plasmas at
equilibrium or even in shock-produced high-density plasmas with $T_e \!<\! T_i$.
\end{appendix}



\begin{thebibliography}{99}
\bibitem{com_1}   The break-down of the classical collision theory
                  first yields too low than negative Coulomb logarithms
          which translates into too low or negative energy
          transfer rates in the Landau-Spitzer theory.
\end{thebibliography}
\end{document}